\documentclass[12pt]{article}
\usepackage{axodraw,a4wide,epsfig}

\newcommand{\bfm}[1]{\mbox{\boldmath$#1$}}

\def\eightSS{\;\parbox{12mm}{\setlength{\unitlength}{0.2mm}
\begin{picture}(60,30)\thicklines
\put(20,15){\circle{20}}
\put(40,15){\circle{20}}
\put(20,15){\oval(24,24)[tl]}
\put(40,15){\oval(24,24)[tr]}
\put(0,15){\line(1,0){10}}
\put(50,15){\line(1,0){10}}
\end{picture}}\;}
\def\sunsetSS{\;\parbox{12mm}{\setlength{\unitlength}{3mm}
\begin{picture}(4,2)\thicklines
\put(2,1){\circle{2}}
\put(2,1){\oval(1.5,1.5)[r]}
\put(0,1){\line(1,0){4}}
\end{picture}}\;}

\catcode`@=11
\newcount\@tempcntc
\def\@citex[#1]#2{\if@filesw\immediate\write\@auxout{\string\citation{#2}}\fi
  \@tempcnta\z@\@tempcntb\m@ne\def\@citea{}\@cite{\@for\@citeb:=#2\do
    {\@ifundefined
       {b@\@citeb}{\@citeo\@tempcntb\m@ne\@citea\def\@citea{,}{\bf ?}\@warning
       {Citation `\@citeb' on page \thepage \space undefined}}%
    {\setbox\z@\hbox{\global\@tempcntc0\csname b@\@citeb\endcsname\relax}%
     \ifnum\@tempcntc=\z@ \@citeo\@tempcntb\m@ne
       \@citea\def\@citea{,}\hbox{\csname b@\@citeb\endcsname}%
     \else
      \advance\@tempcntb\@ne
      \ifnum\@tempcntb=\@tempcntc
      \else\advance\@tempcntb\m@ne\@citeo
      \@tempcnta\@tempcntc\@tempcntb\@tempcntc\fi\fi}}\@citeo}{#1}}
\def\@citeo{\ifnum\@tempcnta>\@tempcntb\else\@citea\def\@citea{,}%
  \ifnum\@tempcnta=\@tempcntb\the\@tempcnta\else
   {\advance\@tempcnta\@ne\ifnum\@tempcnta=\@tempcntb \else \def\@citea{--}\fi
    \advance\@tempcnta\m@ne\the\@tempcnta\@citea\the\@tempcntb}\fi\fi}
\catcode`@=12

\begin{document}

\title{\vskip-3cm{\baselineskip14pt
\begin{flushleft}
\normalsize BI-TP 2004/38\hfill ISSN 0418-9833\\
\normalsize DESY 04-232\\
\normalsize SFB/CPP-04-67 \\
\normalsize TTP/04-24 \\
\end{flushleft}}
\vskip1.5cm
Two-Loop Static QCD Potential for General Colour State}
\author{\small B.A. Kniehl$^a$, A.A. Penin$^{b,c}$,
Y. Schr\"oder$^d$, V.A. Smirnov$^{a,e}$, M. Steinhauser$^b$\\
{\small\it $^a$ II. Institut f\"ur Theoretische Physik, 
           Universit\"at Hamburg,}\\
{\small\it Luruper Chaussee 149, 22761 Hamburg, Germany}\\
{\small\it $^b$ Institut f{\"u}r Theoretische Teilchenphysik,Universit{\"a}t Karlsruhe,}\\
{\small\it 76128 Karlsruhe, Germany}\\
{\small\it $^c$ Institute for Nuclear Research, 
           Russian Academy of Sciences,}\\
{\small\it 60th October Anniversary Prospect 7a, 117312 Moscow, Russia}\\
{\small\it $^d$ Fakult\"at f\"ur Physik, Universit\"at Bielefeld,}\\
{\small\it 33501 Bielefeld, Germany}\\
{\small\it $^e$ Institute for Nuclear Physics, Moscow State University,}\\
{\small\it 119992 Moscow, Russia}}

\date{}

\maketitle

\thispagestyle{empty}

\begin{abstract}
In this letter, we extend the known results for the QCD potential between a
static quark and its antiquark by computing the two-loop corrections
to the colour-octet state. 
\medskip

\end{abstract}


The QCD potential between a static quark and its 
antiquark has for a long time been used
as a probe of the fundamental properties of the strong interactions such as
asymptotic freedom and confinement \cite{Sus}.  Historically, the
potential for a quark-antiquark pair in the colour-singlet state
attracted the most attention because it is a basic ingredient in the theory
of heavy quarkonium and, therefore, of primary phenomenological
interest.  Nowadays, however, there is growing interest in its
colour-octet counterpart. The latter naturally appears in
effective-theory calculations of high-order corrections to the
heavy-quarkonium 
spectrum and decay rates through the so-called ultrasoft contribution
\cite{KniPen}.  Moreover, it determines the properties of glueballinos
and is necessary for the analysis of gluino-antigluino
threshold production \cite{BalPin,KueZer}.  
It is also used in lattice QCD for studying
the behavior of strong interactions at long distances and the interplay
between perturbative and non-perturbative physics \cite{BalPin}.  
This requires knowledge of the corresponding
perturbative corrections which, in contrast to the colour-singlet case,
are not available beyond one loop.  In the present letter, we fill this
gap and compute the ${\cal O}(\alpha_s^2)$ correction to the colour-octet 
static potential.

The perturbative expansion of the colour-singlet potential reads 
\begin{eqnarray}
  V(|{\bfm q}|)&=&-{4\pi C_F\alpha_s(|{\bfm q}|)\over{\bfm q}^2}
  \Bigg[1+{\alpha_s(|{\bfm q}|)\over 4\pi}a_1
    +\left({\alpha_s(|{\bfm q}|)\over 4\pi}\right)^2a_2
    \nonumber\\&&\mbox{}
    +\left({\alpha_s(|{\bfm q}|)\over 4\pi}\right)^3
    \left(a_3+ 8\pi^2 C_A^3\ln{\mu^2\over{\bfm q}^2}\right)
    +\cdots\Bigg],
  \label{singlet}
\end{eqnarray}
where the first term corresponds to the Coulomb potential.
The one-loop coefficient, 
\begin{equation}
  a_1={31\over 9}C_A-{20\over 9}T_Fn_l,
  \label{a1}
\end{equation}
has been known for a long time \cite{Fis,Bil}, while the two-loop coefficient,
$a_2$, has only recently been found \cite{Pet,Sch,KPSS1}.
In Ref.~\cite{KPSS1}, the result of Ref.~\cite{Sch} was confirmed,
\begin{eqnarray}
  a_2&=&
  \left[{4343\over162}+4\pi^2-{\pi^4\over4}+{22\over3}\zeta(3)\right]C_A^2
  -\left[{1798\over81}+{56\over3}\zeta(3)\right]C_AT_Fn_l
  \nonumber\\
  &&{}-\left[{55\over3}-16\zeta(3)\right]C_FT_Fn_l
  +\left({20\over9}T_Fn_l\right)^2,
  \label{a2}
\end{eqnarray}
where $\zeta$ is Riemann's zeta function, with value
$\zeta(3)=1.202057\ldots$.
Here, $C_A=N$ and $C_F=(N^2-1)/(2N)$ 
are the eigenvalues of the quadratic Casimir
operators of the adjoint and fundamental representations of the 
$SU(N)$ colour gauge group, respectively,
$T_F=1/2$ is the index of the fundamental representation, and $n_l$ is
the number of light-quark flavours.
The modified minimal-subtraction ($\overline{\rm MS}$) scheme for the
renormalization of $\alpha_s$ is implied.
The logarithmic term of ${\cal O}(\alpha_s^3)$ in Eq.~(\ref{singlet})
reflects the infrared divergence of the static potential \cite{ADM}.
The particular form of the logarithmic term corresponds to dimensional 
regularization \cite{KPSS2}.
The corresponding infrared-divergent term 
is cancelled against the ultraviolet-divergent one of the
ultra-soft contribution \cite{KniPen} in the calculation
of the physical heavy-quarkonium spectrum \cite{KPSS2,PenSte}.
The non-logarithmic third-order term, $a_3$, is still unknown.

The perturbative expansion of the potential  for the colour-octet
state can be cast in the form
\begin{eqnarray}
  V^o(|{\bfm q}|)&=&
  {4\pi\alpha_s(|{\bfm q}|)\over{\bfm q}^2}
  \left( \frac{C_A}{2}-C_F \right)
  \Bigg[1+{\alpha_s(|{\bfm q}|)\over 4\pi}a^o_1
  +\left({\alpha_s(|{\bfm q}|)\over 4\pi}\right)^2a^o_2
  \nonumber\\&&\mbox{}
  +\left({\alpha_s(|{\bfm q}|)\over 4\pi}\right)^3
  \left(a_3^o + 8\pi^2 C_A^3\ln{\mu^2\over{\bfm q}^2}\right)
  +\cdots
  \Bigg],
  \label{octet}
\end{eqnarray}
where the one-loop coefficient is the same as in the colour-singlet case,
$a_1^o=a_1$. The two-loop coefficient, however, 
differs by a finite renormalization-independent term,
\begin{eqnarray}
  a_2^o&=&a_2+\delta a_2. 
  \label{a2octet}
\end{eqnarray}
Our result is
\begin{eqnarray}
  \delta a_2 &=& C_A^2 \frac{3d-11}{d-5} \left[
  \eightSS
  - \frac{3(d-4)(d-1)}{d-5} 
  \sunsetSS
  \right]
  \nonumber\\
  & =& \left(\pi^4-12\pi^2\right)C_A^2 + {\cal O}(d-4)
  \label{deltaa2}
  \,,
\end{eqnarray}
where $d$ is the space-time dimension.
The non-logarithmic part of the three-loop coefficient, $a_3^o$, is still
unknown.
It is instructive to look at the numerical size of the corrections.
For $N=3$ one obtains $\delta a_2=-189.2$. At the same time, we have
$a_2=155.8 (211.1,268.8)$ and $a_1=4.778 (5.889,7.000)$ for $n_l=5 (4,3)$.
Thus, in the colour-octet case, the two-loop correction is significantly
smaller than for the colour-singlet configuration. 
Depending on $n_l$, it even changes sign.

\bigskip

In the remaining part of this letter, we wish to describe two independent ways
that have been used to evaluate $\delta a_2$.
The first method proceeds along the lines of the analysis
\cite{KPSS1,KPSS2} based on the threshold expansion \cite{BenSmi}.
In general, the threshold expansion
is the proper framework for performing calculations involving
a heavy quark-antiquark
system. It provides rigorous power-counting rules and
natural definitions of the formal expressions obtained 
in the perturbative analysis of the nonrelativistic effective theory.
The corrections to the static potential only arise from the soft
regions of the loop integrals, which are characterized by 
the following scaling of the loop momenta:
$l_0\sim |{\bfm l}|\sim |{\bfm q}|$.
Thus, the calculation of the coefficients $a_i$ and $a^o_i$ 
can be performed in the static limit of NRQCD, $m_q\to\infty$.

Due to the exponentiation of the static potential \cite{Fis}, the
coefficients $a_i$ of the colour-singlet state only
receive contributions from the maximally non-Abelian parts,
leaving aside the terms involving $n_l$.
The selection of these parts
effectively retains the contributions of the soft region,
as the appearance of the
Abelian colour factor $C_F$ indicates the presence of a Coulomb
pinch and thus implies that at least one loop momentum is potential.
The latter contributions just represent iterations of the lower-order potential 
and, therefore, should be excluded from the potential itself.
In the nonrelativistic effective theory, these iterations 
are taken into account in the perturbative solution of the Schr\"odinger
equation about the Coulomb approximation. These 
contributions refer to dynamical rather than
static heavy-quark and -antiquark fields, and the 
Coulomb pinch singularities we encounter in the 
static-limit calculations are resolved by keeping a finite
mass in the nonrelativistic heavy-quark propagator.

The analysis of the colour-octet state is more involved, since, 
in this case, the Coulomb pinches come with all possible
colour factors and cannot be removed by selecting
the maximum non-Abelian ones. Thus, the separation of the
Coulomb pinches should be performed explicitly.
They appear in the Feynman diagrams involving the product of the
nonrelativistic quark and antiquark propagators,
\begin{eqnarray}
\frac{1}{k_0-{\bfm k}^2/(2m_q)+i\varepsilon}\,
\frac{1}{k_0+{\bfm k}^2/(2m_q)-i\varepsilon}.
\end{eqnarray}
In this case, after expanding the quark propagator in 
$1/m_q$, one obtains ill-defined
products like
\begin{eqnarray}
{1\over(k_0+i\varepsilon)^m}\,{1\over(k_0-i\varepsilon)^n}.
\label{pinch}
\end{eqnarray}
Thus, separating the soft and potential regions is 
unavoidable.\footnote{Note that, 
for the diagrams without Coulomb pinches, the separation of the soft 
and potential regions is ambiguous and even gauge dependent.
In such diagrams, the nonrelativistic quark and antiquark propagators
can be safely expanded in $1/m_q$.}
In the soft region, the pole contributions of the quark and antiquark
propagators have to be excluded, and the product in Eq.~(\ref{pinch})
should actually be defined to be
its principal value,
\begin{eqnarray}
{1\over 2}\left[{1\over(k_0+i\varepsilon)^{m+n}}
+{1\over(k_0-i\varepsilon)^{m+n}}\right].
\label{pv}
\end{eqnarray}
In the potential region, the quark and antiquark propagator poles produce
contributions of the form
\begin{eqnarray}
-i\pi\,{m_q\over{\bfm k}^2-i\varepsilon}
\left[\delta\left(k_0-{{\bfm k}^2\over2m_q}\right)
+\delta\left(k_0+{{\bfm k}^2\over2m_q}\right)\right],
\label{pole}
\end{eqnarray}
where the $1/v$ Coulomb singularity shows up explicitly.
After integration over $k_0$, Eq.~(\ref{pole}) yields the nonrelativistic
Green function of the free Schr\"odinger equation. Only 
Eq.~(\ref{pv}) should be taken into account in the calculation of 
the static potential.

At one-loop, there is only one diagram involving a Coulomb pinch, namely, the
planar box, which has the colour factor $C_F^2$ for the colour-singlet state.
Picking up the soft contribution, i.e., using the principal-value
prescription of Eq.~(\ref{pv}) to define Eq.~(\ref{pinch}), we find
the planar box to cancel  
the $C_F^2$ part of the non-planar box, which in total is
proportional to $C_F^2-C_FC_A/2$. This 
explicitly demonstrates the exponentiation of the 
one-loop colour-singlet static potential in momentum space.
However, we can also turn things around and express
the planar box with Coulomb pinches
through the well-defined non-planar box
by actually requiring the cancellation of the $C_F^2$ terms in the sum of all
one-loop diagrams, as is dictated by the exponentiation.
The result for $a_1^o$ as given above is then obtained by simply
replacing the colour-singlet colour factor by the colour-octet one.

\begin{figure}[tb]
  \vspace{2em}
  \begin{center}
    \begin{tabular}{ccccc}
      \begin{picture}(100,60)(0,0)
        \ArrowLine(0,0)(50,0)
        \ArrowLine(50,60)(0,60)
        \ArrowLine(50,0)(100,0)
        \ArrowLine(100,60)(50,60)
        \Gluon(10,0)(10,60){3}{10.5}
        \Gluon(50,0)(50,60){3}{10.5}
        \Gluon(90,0)(90,60){3}{10.5}
      \end{picture}
      &&
      \begin{picture}(100,60)(0,0)
        \ArrowLine(0,0)(50,0)
        \ArrowLine(50,60)(0,60)
        \ArrowLine(50,0)(100,0)
        \ArrowLine(100,60)(50,60)
        \Gluon(10,0)(10,60){3}{10.5}
        \Gluon(50,0)(90,60){3}{12.5}
        \Gluon(90,0)(50,60){3}{12.5}
      \end{picture}
      &&
      \begin{picture}(100,60)(0,0)
        \ArrowLine(0,0)(50,0)
        \ArrowLine(50,60)(0,60)
        \ArrowLine(50,0)(100,0)
        \ArrowLine(100,60)(50,60)
        \Gluon(10,0)(90,60){3}{16.5}
        \Gluon(10,60)(50,0){3}{12.5}
        \Gluon(50,60)(90,0){3}{12.5}
       \end{picture}
      \\ \\
      (a) && (b) && (c)
      \\ \\
      \begin{picture}(100,60)(0,0)
        \ArrowLine(0,0)(50,0)
        \ArrowLine(50,60)(0,60)
        \ArrowLine(50,0)(100,0)
        \ArrowLine(100,60)(50,60)
        \Gluon(10,0)(90,60){3}{16.5}
        \Gluon(50,0)(50,60){3}{10.5}
        \Gluon(90,0)(10,60){3}{16.5}
      \end{picture}
      &&
      \begin{picture}(100,60)(0,0)
        \ArrowLine(0,0)(100,0)
        \ArrowLine(100,60)(0,60)
        \Gluon(20,30)(80,30){3}{10.5}
        \Gluon(20,0)(20,60){3}{10.5}
        \Gluon(80,60)(80,0){3}{10.5}
      \end{picture}
      &&
      \\ \\
      (d) && (e) &&
      \\ \\
    \end{tabular}
    \caption{\small Two-loop Feynman diagrams with ((a) and (b)) and without 
        ((c), (d), and (e)) Coulomb pinches that contribute to $\delta a_2$.}
    \label{fig:one}
  \end{center}
\end{figure}
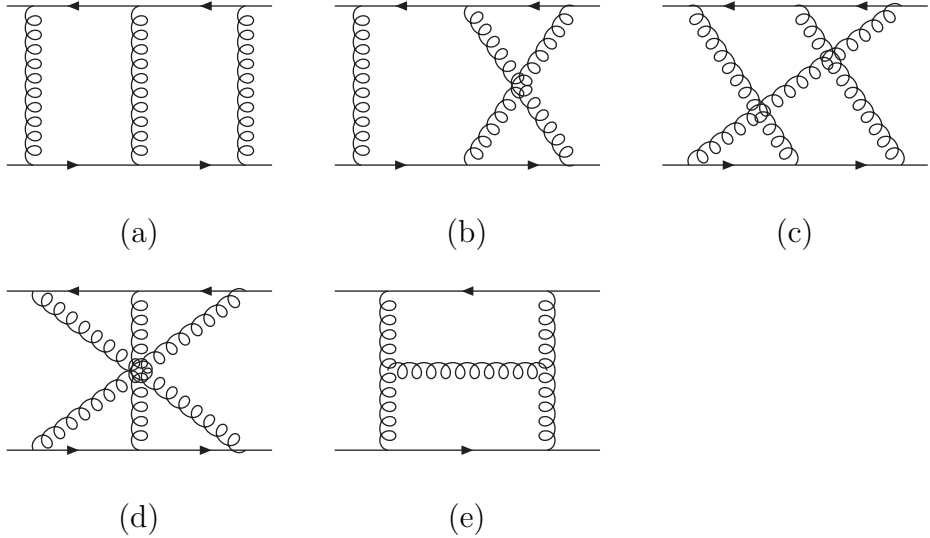

This strategy carries over to two loops. 
Here, we have diagrams with zero, one, or two Coulomb pinches.
For the diagrams without Coulomb pinch, the contribution
to $a_2^o$ is obtained by adopting the correct colour factor.
We divide the Feynman diagrams with Coulomb pinches into those that have
two quark and two antiquark propagators (cf.\ Fig.~\ref{fig:one}) and the
rest. The latter ones are treated directly using the principal-value
prescription of Eq.~(\ref{pv}).
For the former, however, it is simpler to use the exponentiation,
which requires that the diagrams contributing to the
colour factors $C_A C_F^2$ and $C_F^3$ sum up to zero in the colour-singlet
case.
This leads to two equations for the diagrams suffering from Coulomb pinches,
namely, those shown in Figs.~\ref{fig:one}a and b, which can be solved.
This provides a result in terms of the diagrams in Figs.~\ref{fig:one}c and d,
which are free of pinches.  After adopting the colour factors
corresponding to  the colour-octet configuration,
one obtains the  contributions to the
results given in Eqs.~(\ref{a2octet}) and (\ref{deltaa2}).
We wish to mention that the calculation was performed in the general covariant
gauge and that the dependence on the gauge parameter was found to cancel out
in the final result. 

\bigskip

The second method to compute $V^o$ proceeds along the lines of Ref.~\cite{Sch}.
While in the above, we had to assume exponentiation of the 
colour-singlet potential, we will now relax that assumption. 
The reason is that, although exponentiation is plausible
to all orders of perturbation theory, the proof given in 
Ref.~\cite{Fis} holds for the singlet potential in {\em Abelian}
theories only. 

As a starting point, we now expand the logarithm of the $(T\times R)$ 
Wilson loop spanned by the static quark-antiquark pair at distance
$R$ through ${\cal O}(\alpha_s^3)$. 
Taking the limit $T\rightarrow\infty$ (which, in a diagrammatic sense,
`cuts' the Wilson loop twice and restores translational invariance
in the temporal direction, hence guaranteeing energy conservation
at the vertices and leading to simple momentum-space Feynman rules) 
and inserting $SU(N)$ generators
$T^a$ into the purely spatial Wilson lines to obtain the colour-octet
potential to this order (for a manifestly gauge-invariant definition,
see Ref.~\cite{Soto:gi}), we now explicitly keep disconnected as well
as one-particle-reducible diagrams in our expansion. 

At this point, the general structure of the expansion involves
(products of) up to two-loop four-point functions of static quarks
(cf.\ Fig.~\ref{fig:one}).
After Fourier transforming to momentum space, we can choose a special 
point to evaluate these four-point functions, since the potential,
of course, only knows about the distance $R$ of the $q\bar q$ pair,
which in a momentum space representation translates into the momentum 
transfer $|{\bfm q}|$ between the upper and lower lines in Fig.~\ref{fig:one}. 
Hence, effectively, we have to compute two-point functions with
external static quarks, external momentum $q=(0,{\bfm q})$,
and internal static quarks, gluons, ghosts and light quarks, 
with the additional
occurrence of a static (anti-) quark -- gluon two-point vertex,
resulting from the special kinematics.

After performing the colour algebra and exploiting symmetries of the
integrals occurring in the expansion, all integrals which might give
rise to pinch singularities, and had to be treated with caution
in our first approach, cancel exactly. 
Thus, we are left with the task of
computing a class of two-loop two-point integrals
for which there exists a generic algorithm \cite{Sch,SmiSte}, based on
integration by parts (IBP) \cite{IBP}.
The implementation in Ref.~\cite{SmiSte} (see also Chapter~6
of Ref.~\cite{V-book}) is based on Ref.~\cite{Baikov}.

Having generated the relevant set of diagrams and reduced
the occurring Feynman integrals to the set of two-point functions
described above, we now employ the reduction algorithm, which maps
them to a (small) set of so-called master integrals, multiplied
by rational functions in the dimension $d$.
At this stage, we observe cancellation of the gauge-parameter 
dependence, serving as a check for the reduction.
As an additional strong check, we use our implementation \cite{YSibp}
of the strategy to solve a truncated set of IBP relations,
based on lexicographic ordering of integrals \cite{Laporta}.

The set of (massless, two-point) master integrals is known 
analytically in terms of Gamma functions, for generic dimension $d$,
as given, e.g., in Ref.~\cite{Sch}. 
Expanding prefactors as well as master integrals about $d=4-2\epsilon$ 
and renormalizing the gauge coupling, we again arrive at 
Eq.~(\ref{deltaa2}).

\bigskip

To conclude, we have evaluated the ${\cal O}(\alpha_s^2)$ correction 
to the colour-octet static potential using two independent techniques.
Both evaluations are in agreement, giving us confidence in our 
main result, Eq.~(\ref{deltaa2}).

\bigskip

\noindent
{\bf Acknowledgements:} 
Y.S. would like to thank the phenomenology 
group at Hamburg University for hospitality.
The work of A.A.P. was supported in part by BMBF Grant No.\ 05HT4VKA/3.
The work of V.A.S. was supported in part by RFBR Project
No. 03-02-17177, Volkswagen Foundation Contract No. I/77788, and DFG
Mercator Visiting Professorship No. Ha 202/1.
This work was supported by BMBF Grant No. 05HT4-GUA/4,
HGF NG-VH-008, and SFB Grant No. TR 9.

\end{document}